# Data cluster analysis and machine learning for classification of twisted bilayer graphene

Tom Vincent[1,2], Kenji Kawahara[3], Vladimir Antonov[4,2], Hiroki Ago[3] and Olga Kazakova[1,*]
1. National Physical Laboratory, Hampton Road, Teddington TW11 0LW, UK
2. Department of Physics, Royal Holloway University of London, Egham TW20 0EX, UK
3. Global Innovation Center (GIC), Kyushu University, Fukuoka 816-8580, Japan
4. Skolkovo Institute of Science and Technology, Moscow, 143026, Russia

Email: olga.kazakova@npl.co.uk

Twisted bilayer graphene (TBLG) has emerged as an exciting new material with tunable electronic properties ranging from superconductivity to correlated insulating phases. But current methods of fabrication and identification of TBLG are painstaking and laborious. In this work, we combine Raman spectroscopy with the Gaussian mixture model (GMM) data clustering algorithm to identify areas with particular twist angles, from a TBLG sample with a mixture of orientations. We present two approaches to this cluster analysis: training the GMM on Raman parameters returned by peak fits, and on full Raman spectra with dimensionality reduced by principal component analysis. In both cases we demonstrate that GMM can identify regions of distinct twist angle from within Raman datacubes. We also show that once a model has been trained, and the identified clusters labelled, the model can be reapplied to new Raman scans to assess the similarity between the materials in the new region and the testing region. This could enable high-throughput fabrication of TBLG, by allowing computerised detection of particular twist angles from automated large-area scans.

## INTRODUCTION

The demonstration of superconductivity in twisted bilayer graphene (TBLG)[1] sparked a flurry of new research into this material[2–9]. It is formed of two sheets of single layer graphene (SLG) stacked together, with one lattice rotated relative to the other. Interference between the two lattices adds an extra moiré superlattice potential, whose period depends on the twist angle. This modifies the electronic bandstructure of the bilayer graphene (BLG), leading to novel properties such as enhanced photocurrent generation[10], superconductivity[1,2] and correlated insulating phases[9,11] at particular twist angles. With the advent of twist engineering, bilayer graphene has expanded from a single material to a whole family of materials with continuously tunable properties. More recently, twist engineering has been applied to bilayers and heterobilayers of other 2D materials, demonstrating its wider applicability beyond graphene[12–15].

However fabricating TBLG samples is a highly time and labour-consuming process. The most common method used is the 'tear and stack' technique[1,2,16], which involves manually picking up and rotating flakes before depositing them. These samples have a tendency to relax into twist angles other than those intended[1], due to the varying interlayer potential associated with different twist, so methods to accurately determine the twist angle are vital. Existing methods, including Raman spectroscopy[17–20], transmission electron microscopy[18] and low energy electron diffraction[20], are inherently laborious and need highly-specialised human input for data interpretation.

In this work, we explore new analysis techniques for Raman spectroscopy data acquired from a TBLG sample with a wide range of different twist angles. Firstly, we present a manual analysis to identify features in the Raman spectra associated with particular twist angles. We then apply a data clustering technique, the Gaussian mixture model. We demonstrate a priori application of the technique to identify distinct material types within Raman maps, showing that it can identify the same key regions found by human analysis, and that it can provide further statistical information on features such as strain in the graphene. We then show that once a model has been trained on an area of interest, it can be used to successfully categorise new regions of TBLG in terms of similarity to desired twist angles. We present two approaches to dimensionality reduction of the Raman spectra: first using parameters returned by fitting peaks to the data, and then using principal component analysis. By doing this, we demonstrate that data clustering has the potential to enable high-throughput classification of TBLG, by allowing computerised detection of particular TBLG species from automated large-area scans.

## RESULTS AND DISCUSSION

### Sample characteristics

The TBLG sample used in this work consists of two batches of graphene prepared by chemical vapour deposition (CVD), which were transferred to the same $SiO_2$ substrate, as shown in **Fig. 1a**. Batch 1 is formed of hexagonal, single-layer grains on the order of tens of microns in lateral size, with a stochastic distribution of orientations. Batch 2 is formed of larger single-layer grains on the order of hundreds of microns in lateral size, with a single dominant alignment. Batch 1 sits on top of batch 2, and the areas where the two batches overlap form TBLG, with a stochastic distribution of twist angles. See **Methods** for more detail on the fabrication process.

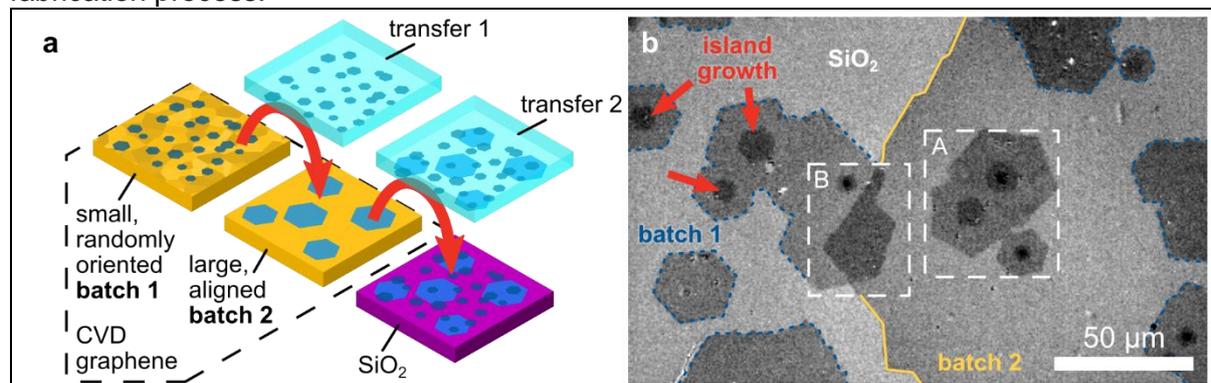

**Fig. 1 Sample details. (a)** Preparation process. Randomly oriented (batch 1) and aligned (batch 2) graphene grains are grown on polycrystalline Cu foil and Cu(111) film respectively, then transferred to the same $SiO_2$ substrate. **(b)** Optical microscope image showing resulting sample, with areas of TBLG formed by overlapping graphene batches. Regions A and B, focused on in the rest of this work, are indicated with white rectangles. Randomly oriented grains from batch 1 are outlined with blue dotted lines, a large grain from batch 2 is outlined with a solid yellow line. No outlines are drawn inside regions A and B to show the sample more clearly. Multilayer islands are indicated with red arrows.

An optical microscope image of a representative area of the sample is shown in **Fig. 1b**.

A key advantage of this fabrication process is that we can infer the orientation of the graphene lattice from the angle of the hexagonal grain edges, which gives us a reasonable estimation of the TBLG twist angle. We use the shorthand TBLG$_{\theta°}$ to refer to TBLG with an estimated angle $\theta°$. The edges are not perfectly straight, which leads to an uncertainty on the order of ±1°.

The grains of batch 1 feature several small patches of bilayer and multilayer graphene. These appear as darker areas in the optical image, towards the centre of some grains, and are highlighted by red arrows in **Fig. 1b**. We shall refer to these patches as multilayer islands throughout this work. We stress that any bilayers which result from these islands are produced by a different mechanism to the artificially created bilayers formed by overlapping grains from batches 1 and 2, and as such we should not expect the same varied distribution of twist angles. By comparison, the grains in batch 2 are almost entirely free of multilayers.

In the rest of this work, we focus on two similar regions of the sample. Region A is used for training the data clustering models, and region B is used to test the pretrained models. The locations of these regions are shown in **Fig. 1b**.

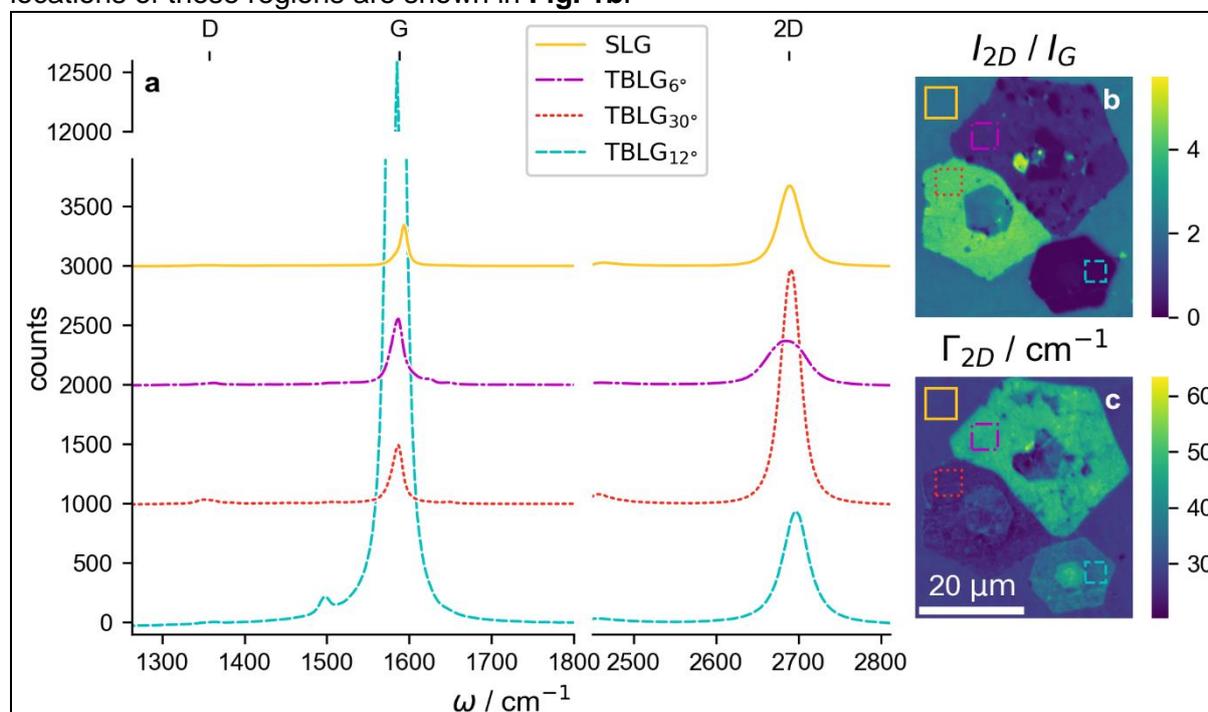

**Fig. 2 Raman features of twisted bilayer graphene: region A. (a)** Representative averaged Raman spectra, showing large variation with twist angle. The averages were taken from within the squares of the corresponding colour and line style shown in **b** and **c**. **(b, c)** Maps of the 2D-to-G peak intensity ratio and 2D peak FWHM.

**Raman analysis**

Before we discuss data clustering, we first present a more traditional analysis, to explain the main Raman features of this TBLG sample. A Raman datacube was collected from region A. **Fig. 2a** shows a set of representative Raman spectra taken from key areas. The dominant features of the graphene Raman spectrum are the G and 2D peaks, whose positions are shown at the top of **Fig. 2a**. We will analyse these peaks in terms of their intensity, Raman shift and full width at half maximum (FWHM), which we denote by $I_P$, $\omega_P$ and $\Gamma_P$, where $P$ denotes the peak in question. **Figs. 2b** and **c** show spatial maps of the 2D-to-G peak intensity ratio and the 2D peak FWHM. These are typical metrics used to indicate the quality and layer number of graphene[21]. The locations of the displayed spectra are indicated in these maps.

The spectrum from the single layer graphene (SLG) has a 2D-to-G ratio of ~2, which is typical for monolayers on $SiO_2$. The graphene D peak, whose position is also indicated in **Fig. 2a**, and whose intensity correlates with defect density, is notably absent. This confirms that the graphene from CVD batch 1 is high quality and single layer, as designed.

There are three TBLG grains in region A, which can be seen clearly in **Figs. 2b** and **c**. As discussed above, we estimated the twist angles from the grain edge orientations to be ~6°, ~12° and ~30° using the optical image in **Fig. 1b**. **Fig. 2a** shows that all three domains have a small D peak, which indicates that batch 2 has a slightly higher defect density than batch 1, though the D-to-G intensity ratio is generally low at <0.1.

The $TBLG_{6°}$ spectrum has a broadened 2D peak and a 2D-to-G ratio of <1. This is typical for graphene with a twist angle below ~10° [ref. [17]].

The $TBLG_{30°}$ spectrum has an enhanced 2D peak, and appears similar to the expected spectrum from SLG, which also matches predictions and observations for large twists from literature[17–19].

The $TBLG_{12°}$ spectrum has a strong (~30×) enhancement of the G peak. This is caused by van Hove singularities (saddle points) in the superlattice bandstructure with a separation close to the energy of our excitation laser[17,18,20], which leads to a strongly resonant Raman process. For a 532-nm excitation laser, we calculate that this enhancement should occur for TBLG with a twist close to 11.9° [ref. [17]], which is consistent with our estimated angle.

The maps also show that the multilayer islands at the centre of each grain have distinct Raman characteristics, and that some of them contain grain boundaries themselves.

A typical Raman analysis may involve comparing the parameters of many peaks, as demonstrated here. This can be a laborious process, which needs expert input, and often requires fitting many peaks to the collected data. A typical set of maps showing the results of fitting to the 2D, G and D peak from region A is shown in **Supplementary Fig. 1**.

**Gaussian mixture model**

GMM is a data clustering technique which assumes a dataset is drawn from a set of $N$ normally distributed clusters. It uses the expectation-maximisation algorithm[22] to find the means, covariance matrices and weightings of the $N$ Gaussian probability distributions that best describe the data.

The algorithm compares the distance between points in an $n$-dimensional space, where each dimension is formed by a separate observation (*e.g.* $\omega_{2D}$, $I_G$). However, the observations may have different units or variances, which could give unfair weighting to some observations over others. We therefore use Mahalanobis normalisation[23] to enforce that each dimension is unitless, has zero mean and has unit variance. In the rest of this work, for an arbitrary parameter $p$, we use the shorthand $p'$ to denote it's Mahalanobis normalised equivalent.

We use a two-dimensional example to illustrate the GMM fitting process. **Fig. 3a** is a scatterplot showing the normalised positions for the G and 2D peaks from region A. The unnormalised maps of the same parameters are shown in **Figs. 3b** and **c**. A GMM with $N$=8 Gaussian clusters was fit to the data shown, and the points were assigned a colour based on which of the resulting clusters they were most likely to belong to. Numbers next to each cluster serve as labels and are ordered from the most (1) to the least (8) populated cluster. Confidence ellipses for each Gaussian cluster have also been added to **Fig. 3a**, to highlight the shape of the clusters. These are drawn at 2σ away from the mean, where σ is the direction-dependent standard deviation.

In general, the positions and covariances of the identified clusters will depend somewhat on the number of clusters fit to the data. In this case, the value $N=8$ was chosen for the purpose of demonstration, based on the number of distinct colours that could be shown in a visualisation. There are certain techniques, which can infer the number of clusters to use from the data itself[24], however these are beyond the scope of this work. For real-world applications, prior knowledge of the expected number of clusters can often be used to inform the choice of $N$.

The map in the inset of **Fig. 3a** shows each coloured observation in the scatterplot mapped back to its original spatial position. This serves as a key to the colours, and shows that identified clusters correspond to the same distinct material types that we identified with the traditional Raman analysis above. There are clusters which account for all three kinds of TBLG (clusters 2, 3 and 6), as well as the island growth (cluster 4) and SLG (clusters 1 and 5). The width of each cluster is a reflection of the distribution of values within the cluster.

We also note that the GMM has included some wide clusters with a low weighting (clusters 7 and 8), which account for observations that do not fit into any of the main clusters. These can be viewed as a background to the total probability density function. Points that are identified as belonging to these background clusters are few, and often localised at the spatial boundaries between materials, where the Raman spectra may have features of both materials present due to the finite beam size. Including these background-type clusters is possible because GMM explicitly accounts for overlapping distributions. This is a key advantage of GMM over other data clustering techniques such as *k*-means, which assigns observations to a cluster based on the closest cluster centre, and positions the centres to minimise the intra-cluster variance. A *k*-means fit to the same data as **Fig. 3** is shown in **Supplementary Fig. 2** for comparison.

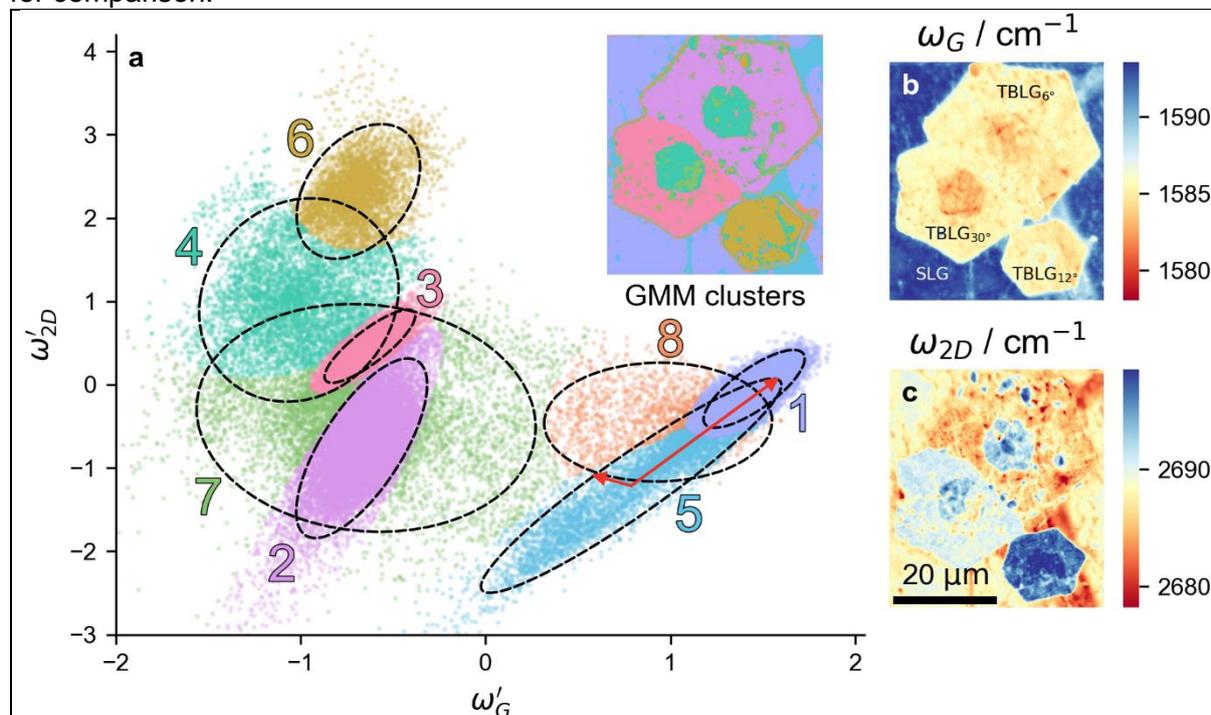

**Fig. 3 Gaussian mixture model. (a)** Scatterplot of Mahalanobis normalised positions of the G and 2D peaks from region A. Points are coloured according to the cluster assigned to

> them by an 8-component GMM fit to the data shown. Black dashed lines show the $2\sigma$ confidence ellipses for each cluster. Red arrows show the principal component vectors (scaled to $2\sigma$) for one cluster. Inset: key to cluster colour and spatial distribution. **(b, c)** Maps of the unnormalised G and 2D peak positions. Twisted grains and single layer areas are indicated in **(b)**.

It is interesting to note that the SLG has been split into two clusters (1 and 5), caused by small areas within the region with downshifted peak positions. We attribute this to variation in substrate interaction, which may induce both strain and doping to the graphene.

In single layer graphene, strain induces a linear shift of the G and 2D positions, whose gradient depends on the type and orientation of the strain[25–27]. We use eigendecomposition of covariance matrices, to find the principal component vectors for the SLG clusters. These vectors show the directions which correlate with most variance within a particular cluster. Red arrows in **Fig. 3a** show the principal components for a single cluster as an example. We can calculate the true gradients of these principal components by transforming them back into the unnormalised data space. If the origin of the peak shifting is solely due to strain, we should expect the gradient of the dominant principal component vector to lie in the range 2.02-2.44 [ref. [25]]. For the highest weighted cluster, corresponding to most SLG (purple), we obtain a gradient of 2.02, but for the cluster corresponding to the downshifted SLG (blue) we obtain a lower gradient of 1.69. This shows that the position shift cannot be explained by strain alone, and there is likely also a component of the variation induced by doping inhomogeneity[25–28]. This is a good example of the analytical power of GMM, beyond simple material identification.

**Classifying new data using a pretrained model**

In this section, we show an example of a GMM trained on one area of the sample (region A) being used to identify similar materials in a new area (region B). **Fig. 4** shows the clusters in region A identified by a three-dimensional GMM with $N$=8, whose inputs are the intensity, position and width of the 2D peak. The maps in **Figs. 4a-h** show the relative likelihood that a point belongs to each of the identified clusters. The positions of the identified clusters in the normalised data space are shown by the colours of the points in the scatterplot in **Fig. 4i**, which are set to match the peak colour of the corresponding maps.

We have assigned a label to each cluster, based on the traditional Raman analysis performed above, and on the twist angle estimated from the optical image in **Fig. 1**. We can see that the SLG is once again split due to strain- and charge-induced effects. We also see that the TBLG$_{6°}$ region has split into two clusters. This may be caused by small domains of locally relaxed twist angle, compared to the global twist of the whole grain[20]. Similarly, we see that there are two clusters corresponding to island growth, which we attribute to domains with AB and non-AB stacking. Where multiple clusters have been assigned to one type of material in this way, we use superscript numbers in parentheses to refer to the individual clusters (*e.g.* SLG$^{(1)}$ and SLG$^{(2)}$, for two distinct SLG clusters).

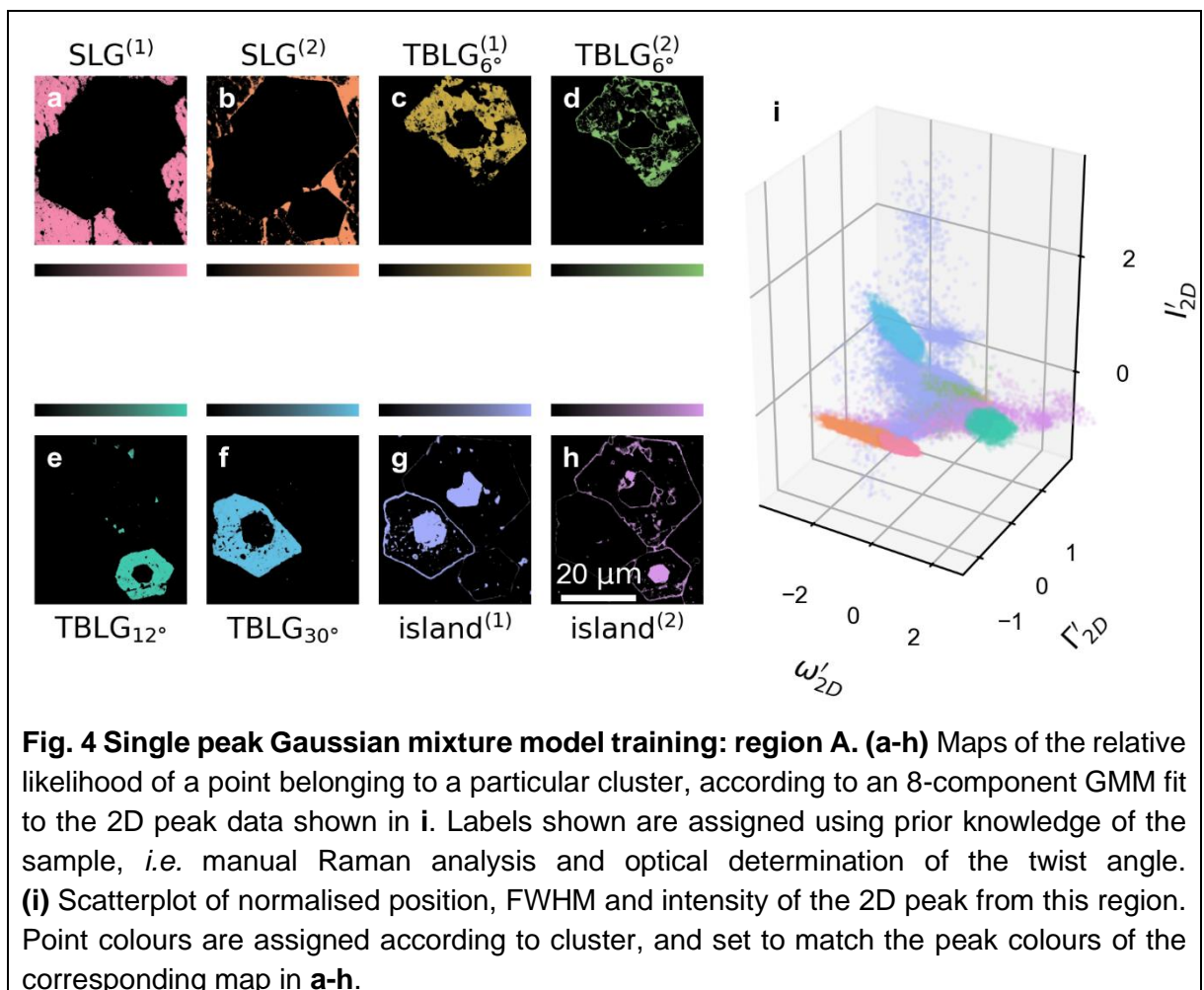

**Fig. 4 Single peak Gaussian mixture model training: region A. (a-h)** Maps of the relative likelihood of a point belonging to a particular cluster, according to an 8-component GMM fit to the 2D peak data shown in **i**. Labels shown are assigned using prior knowledge of the sample, *i.e.* manual Raman analysis and optical determination of the twist angle. **(i)** Scatterplot of normalised position, FWHM and intensity of the 2D peak from this region. Point colours are assigned according to cluster, and set to match the peak colours of the corresponding map in **a-h**.

Once a model has been trained, and the desired materials have been labelled, the clusters can be reapplied to new data, to judge how similar the new material is to the training material. We apply the model trained on region A in **Fig. 4** to classify the data from a Raman datacube collected from region B.

Region B shares many features with region A, including SLG areas, a $TBLG_{12°}$ area, a $TBLG_{30°}$ area and some island growth. However there are also some key differences. The island growth in region B does not overlap with batch 1, so it forms BLG rather than trilayer. Additionally, the $TBLG_{30°}$ is the result of island growth rather than the stacking of two separate graphene batches. There is also a small area of AB stacked BLG, formed by island growth, and no $TBLG_{6°}$ grain. A full set of Raman maps and some spectra from key areas of region B are shown in **Supplementary Figs. 3** and **4**.

The results from classifying Region B (indicated in **Fig. 1b**) with the pretrained GMM are shown in **Fig. 5**. The maps in **Figs. 5a-h** show the relative likelihood that a point belongs to each of the previously identified clusters from region A. The labels and colours are the same as those assigned in **Fig. 4**. The scatterplot in **Fig. 5i** shows the distribution of points in the same normalised space as for **Fig. 4** (*i.e.* each parameter normalised to the mean and variance of region A).

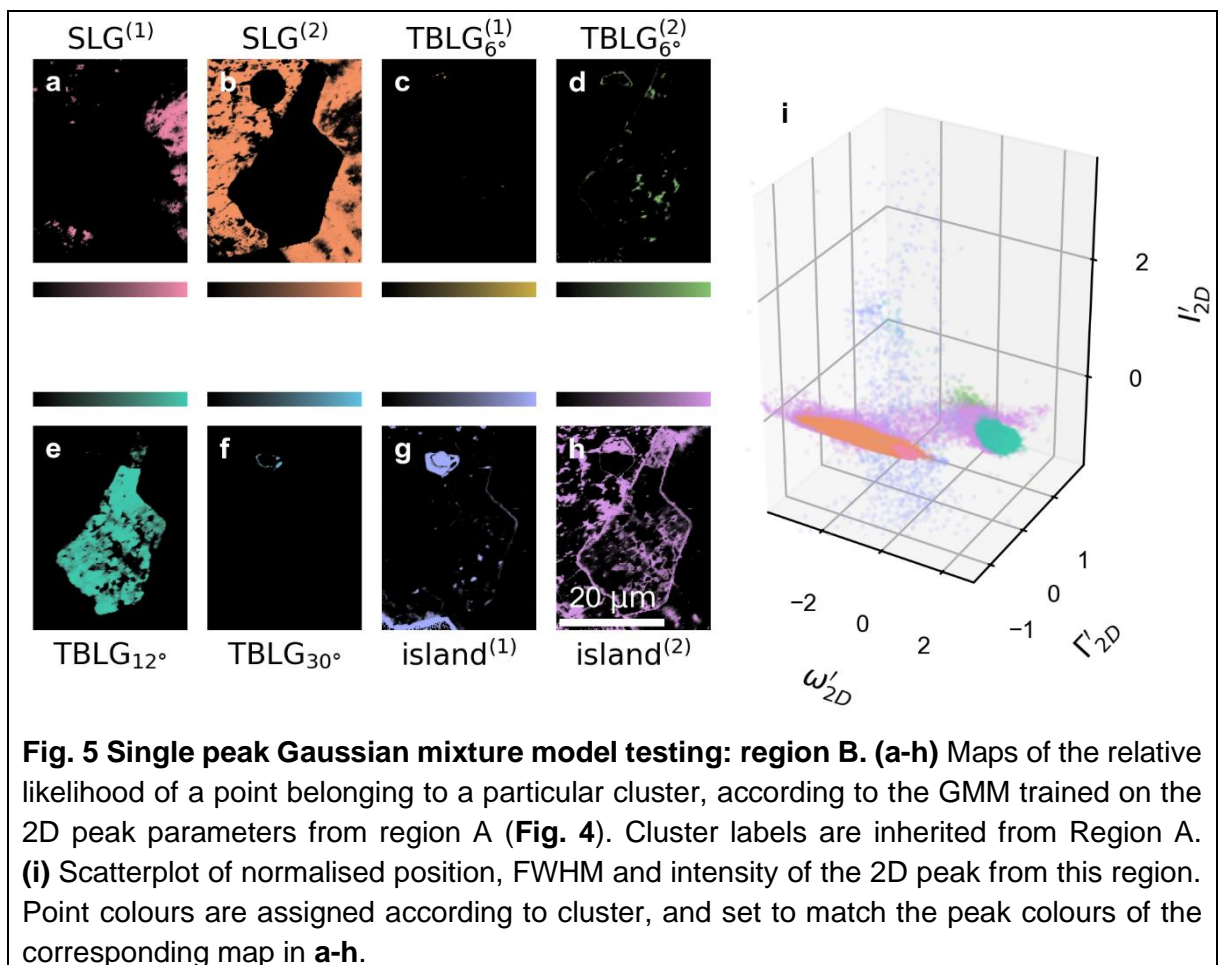

**Fig. 5 Single peak Gaussian mixture model testing: region B. (a-h)** Maps of the relative likelihood of a point belonging to a particular cluster, according to the GMM trained on the 2D peak parameters from region A (**Fig. 4**). Cluster labels are inherited from Region A. **(i)** Scatterplot of normalised position, FWHM and intensity of the 2D peak from this region. Point colours are assigned according to cluster, and set to match the peak colours of the corresponding map in **a-h**.

**Figs. 5a** and **b** show that the GMM has correctly identified the SLG in region B. Interestingly we see a reversal of the distribution of the two clusters from region A. In region B most points belong to the cluster marked SLG$^{(1)}$ (**Fig. 5a**), whereas in region A most points were in the cluster marked SLG$^{(2)}$ (**Fig. 5b**). We attribute this to strain variations between regions A and B. We also note that the SLG in the left of region B is from batch 2, whereas the SLG in the right, and in region A is from batch 1 (see **Fig. 1b**). The cluster marked SLG only appears significantly in the batch 1 SLG. The cluster for the TBLG$_{12°}$, and the island growth clusters have also correctly identified the same features from region B.

TBLG$_{6°}$ is absent from region B so the corresponding maps in **Figs. 5c** and **d** are mostly empty. Although the map showing the TBLG$_{30°}$ cluster in **Fig. 5f** is also mostly empty, this reflects the small amount of TBLG$_{30°}$ present. It should be noted that the TBLG$_{30°}$ in region B is formed by island growth, and the GMM classification identifies most of it as having more similarity to the islands from region A, than to the TBLG$_{30°}$ from region A (compare **Figs. 5f** and **g**).

The example above is a good illustration of one of the pitfalls of this sort of automated data analysis: much care needs to be taken when labelling clusters from the training set. As the fitting is performed without human guidance, the model may not have converged on the features that one would expect. Any error in the labelling is likely to lead to a misidentification of features further down the line.

Overall, we have shown that, once trained and labelled, the GMM is able to successfully identify similar materials from a new region of the sample. The fact that we were able to identify all these regions from the parameters of a single peak also shows that the GMM can enhance the discriminating power of relatively small numbers of parameters. The same process is successful when using the other Raman peaks of graphene, this is shown for the G and D peaks in **Supplementary Figs. 5-8**.

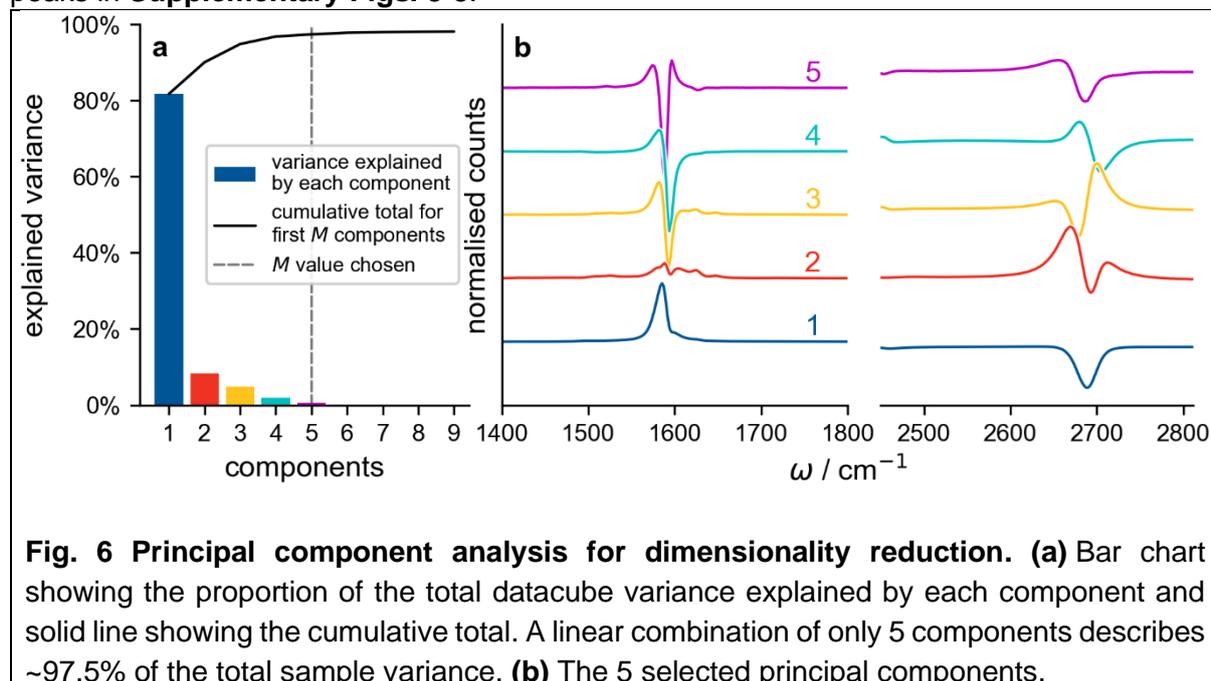

**Fig. 6 Principal component analysis for dimensionality reduction. (a)** Bar chart showing the proportion of the total datacube variance explained by each component and solid line showing the cumulative total. A linear combination of only 5 components describes ~97.5% of the total sample variance. **(b)** The 5 selected principal components.

**Application to full spectra**

The demonstrations above used peak parameters as inputs. This can be useful for analysis but requires fitting of peaks to the spectra, which can be time consuming and often requires expert intervention to optimise the fits. For fast material identification it would be more practical to put the whole Raman spectrum as input into the GMM without any supervised pre-processing. Although this is possible, the procedure is computationally expensive, as there are typically ~1000 points in a typical spectrum, leading to a ~1000-dimensional space for the fitting. This is not ideal as it takes a long time to converge, and has a higher likelihood of converging on spurious local minima.

We circumvent this problem by reducing the dimensionality of our data using principal component analysis (PCA), which works by using eigendecomposition of the covariance matrix of an entire dataset to identify the principal component vectors. For an $M$-dimensional space, there will be $M$ vectors of length $M$. The data can then be expressed as a linear combination of these components. Each of these vectors accounts for a certain proportion of the total variance of the dataset. When there are strong correlations between parameters, as there are between neighbouring points in Raman spectra, we can account for a large fraction of the total variation using only a few components.

We ran PCA on the datacube from region A, normalised such that each spectrum forms a vector with unit magnitude. **Fig. 6a** shows the total variance explained by the 9 most dominant components. It shows that we can reduce the dimensionality of our dataset from 1015 to 5, by expressing our data using only the first 5 components and still explain ~97.5% of the total variance. For Raman data, the principle component vectors resemble spectra. We show the normalised vectors for the 5 selected components in **Fig. 6b**.

A spectrum can then be expressed as a 5-dimensional vector of coefficients, corresponding to the amount of each component present. Spatial maps of these coefficients across both regions A and B are shown in **Supplementary Fig. 9**

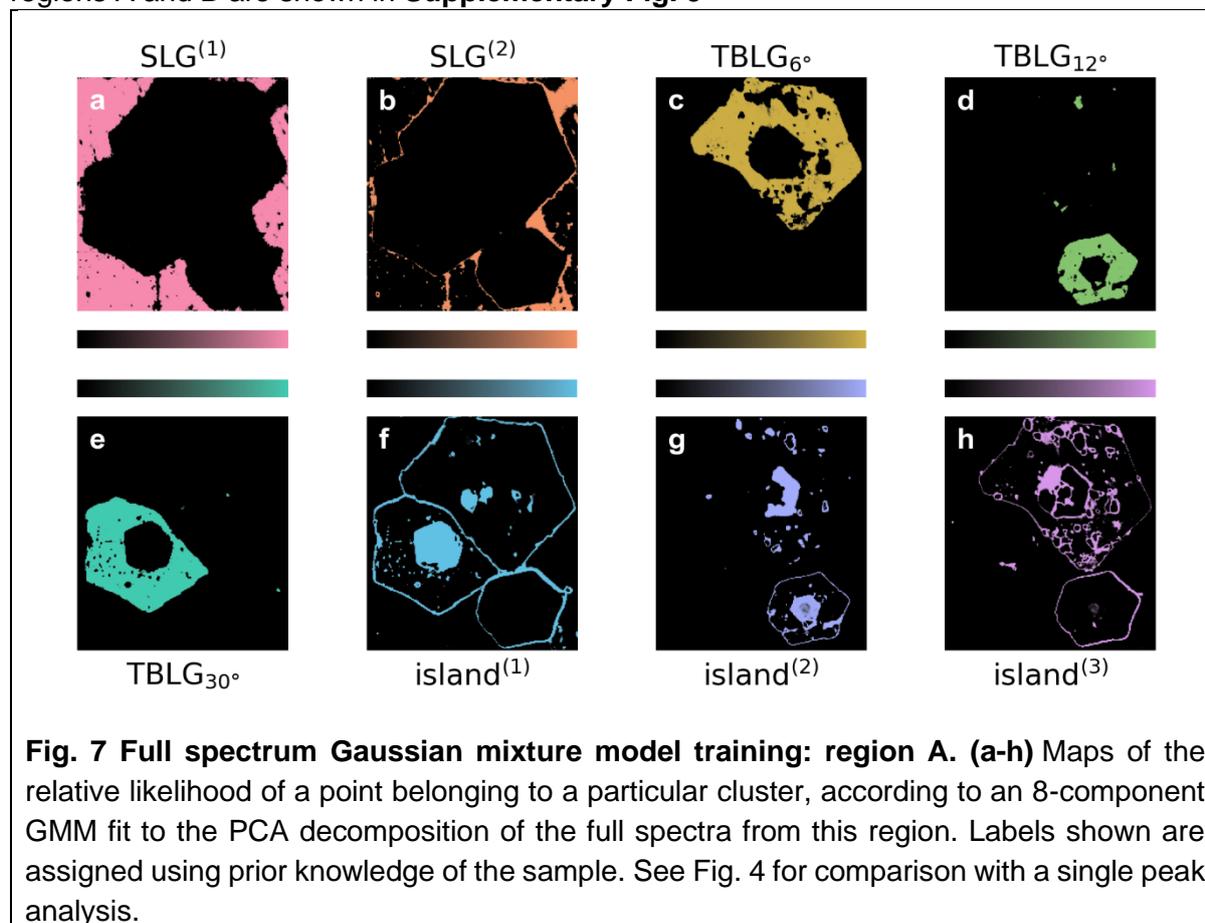

**Fig. 7 Full spectrum Gaussian mixture model training: region A. (a-h)** Maps of the relative likelihood of a point belonging to a particular cluster, according to an 8-component GMM fit to the PCA decomposition of the full spectra from this region. Labels shown are assigned using prior knowledge of the sample. See Fig. 4 for comparison with a single peak analysis.

We use these coefficients as input to a GMM trained on region A with $N$=8, and then assign labels as we did in the preceding section. These are shown in **Fig. 7**. As the input vector is now 5-dimensional we cannot show a scatterplot showing the cluster distribution in data space as we did previously, but we show a pairwise scatterplot of each coefficient in **Supplementary Fig. 10**. We see that the PCA decomposition has once again identified the same key regions we identified above, with the SLG area split into two clusters by varied strain and charge doping, however we also see three distinct island growth clusters, which we attribute to different stacking configurations.

As before, we apply the pretrained model to classify region B, as shown in **Fig. 8**. We see that the SLG, $TBLG_{12°}$ and $TBLG_{30°}$ are correctly identified, and the absent $TBLG_{6°}$ does not appear in the map. There is some notable uncertainty in identification of the island growth regions, which also pick up a lot of the boundaries between regions. However this is shown in both training and test data, so could easily be accounted for with more verbose labelling. This indicates that these clusters are of the wide, low-weighting background-type discussed above. Additionally, these clusters are all composed of features that are not necessarily shared by both regions (there are no trilayer islands in region B), so it is not surprising that the GMM does not find the same features in each region.

This shows that the PCA dimensionality reduction expresses the data well enough to be a viable alternative to curve fitting as a pre-processing step for GMM. Which could potentially reduce the amount of expert time needed to perform this analysis.

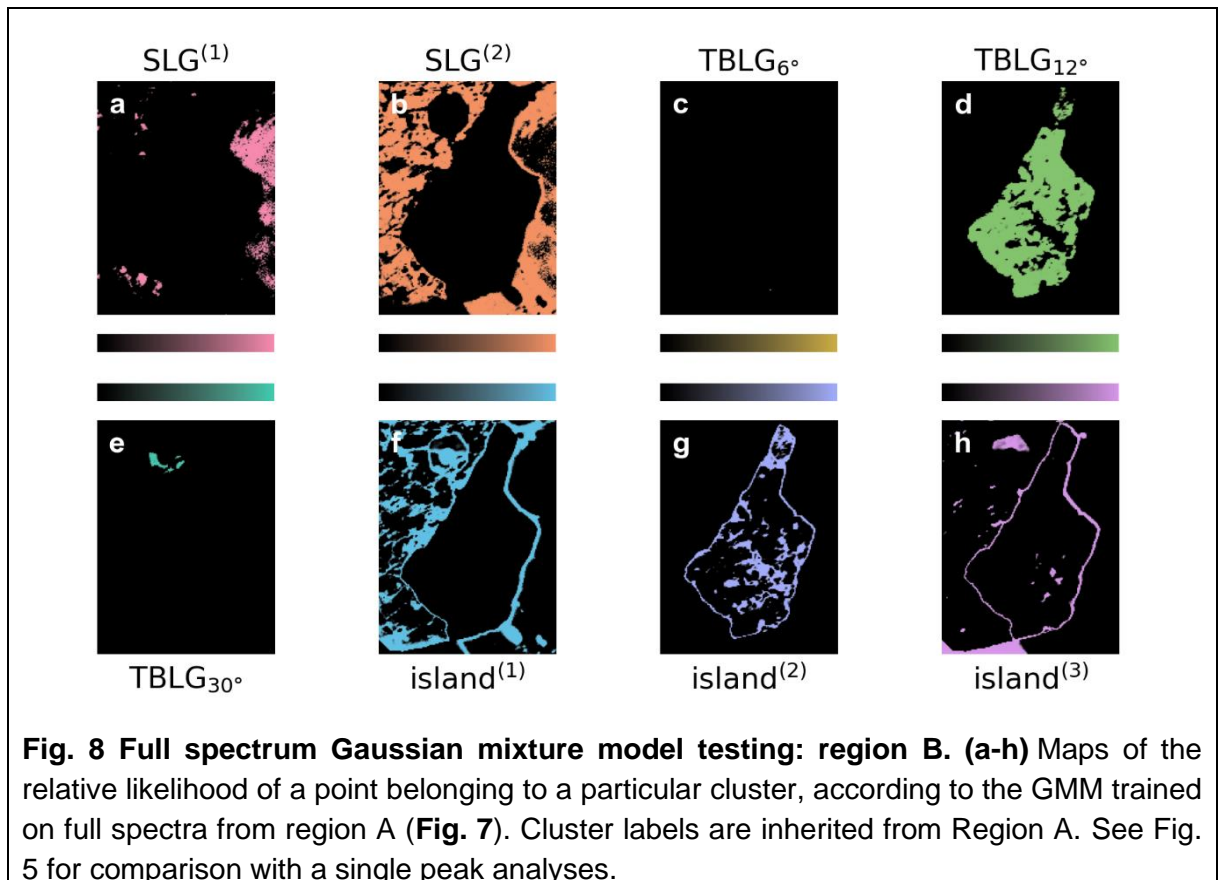

**Fig. 8 Full spectrum Gaussian mixture model testing: region B. (a-h)** Maps of the relative likelihood of a point belonging to a particular cluster, according to the GMM trained on full spectra from region A (**Fig. 7**). Cluster labels are inherited from Region A. See Fig. 5 for comparison with a single peak analyses.

## CONCLUSIONS

In this work, we have demonstrated that, following initial training, GMM data clustering can be used successfully to quickly identify desired species of TBLG from Raman spectra. We showcased two approaches to reduce the dimensionality of the spectra before passing them to the algorithm: fitting peaks to the spectra, and pairing the GMM with PCA.

The advantage of using parameters returned from peak fitting as GMM inputs is that the returned clusters yield easy-to-interpret physical information on the per-cluster parameter distribution. This makes it a good complement to a more traditional Raman analysis approach.

We showed how the statistical information returned by the GMM after training, *i.e.* the per-cluster means and covariance matrices, can be used to reveal important details about the sample. In our example we showed that the shifting of the G and 2D peaks in the SLG, attributed to inhomogeneous substrate interaction, could not result solely from strain, and must also have a contribution from changes in charge doping. This demonstrates a significant advantage of GMM above other clustering algorithms, which do not return the same statistical information.

By contrast, the physical meanings of the cluster distributions from GMM paired with PCA are harder to interpret. However the algorithms can take entire Raman spectra as input, which is a key benefit, as it means that acquired spectra could be assessed immediately, without the need for human intervention. It therefore enables automation of a process which often requires many hours of expert scrutiny, condensing the total analysis time down to seconds.

Like any machine learning tool, the performance of a GMM for classification will depend on the training data used and appropriate labelling of the identified clusters. For example, the areas we showed with the most indeterminacy in classification were the areas of our sample which differed most between our training and testing regions (*e.g.* the TBLG$_{30°}$ formed by island growth in **Figs. 5f** and **g**). In practice this can be mitigated by careful initial labelling, and appropriate selection of training data.

As we have shown above, this tool is very well suited to classifying materials based on Raman spectra, but the techniques shown here can also be applied to any set of colocalised measurements, for example other forms of spectroscopy, or data acquired via scanning probe microscopy. Likewise, we have demonstrated the technique for successfully identifying TBLG, but it is equally suited to identifying species in any type of mixed-material sample, whose materials produce distinct measurements.

Automated analysis, using the machine learning techniques shown here, has potential to dramatically speed up material identification, particularly when paired with automated scanning. This will have profound consequences for the future of material optimisation and device fabrication, as it will remove a major bottleneck in the fabrication process. It can also make samples with a stochastic distribution of materials, such as the TBLG shown in this work, more practicable for inclusion in devices, as automated scanning and analysis could identify and locate target materials from a whole substrate in a matter of hours, a process which previously may have taken days for a trained scientist.

## METHODS

**Sample fabrication**

To produce the TBLG sample used in this work we prepared two batches of graphene on Cu by ambient-pressure chemical vapour deposition (CVD) at 1075 ºC using CH$_4$ feedstock. Different Cu catalysts were used to yield different distributions of grain size and orientation[29]. Batch 1 was grown on polycrystalline Cu foil (Nilaco Co.), resulting in single layer grains on the order of tens of microns lateral size, with a stochastic distribution of orientations. Batch 2 was grown on Cu (111) thin film sputtered on c-plane sapphire[30,31], resulting in large single layer grains on the order of hundreds of microns lateral size, which are aligned with the Cu lattice. The graphene grown on polycrystalline Cu foil contained many multilayer grains, while that grown on Cu(111) was almost free from multilayers.

These two batches were then transferred to a single SiO$_2$ substrate with a polymethyl methacrylate (PMMA) layer, by etching away the Cu in ammonia persulfate (APS) aqueous solution. The PMMA/graphene stack (batch 1) was transferred onto graphene/Cu(111) (batch 2), followed by removal of the PMMA with acetone. Because the PMMA film used in the transfer of the batch 1 graphene had deteriorated during the first transfer step, we removed the old PMMA layer and spin-coated a new PMMA layer for the second transfer process. Finally, the PMMA/graphene (batch 1)/graphene (batch 2) stack was transferred onto a SiO$_2$ substrate. The areas where the two grains overlap form TBLG, with a stochastic distribution of twist angles.

A graphical illustration of the whole fabrication process can be seen in **Supplementary Fig. 12**.

**Raman**

Raman mapping was performed using a Renishaw inVia confocal Raman microscope. We used a 532 nm excitation laser focused through a 100× objective, with a numerical aperture of 0.85. The laser power incident on the sample was ~0.5 mW. A 1800 line mm$^{-1}$ diffraction grating was used. The measurements were taken in high confocality mode, resulting in an estimated spot size of ~450 nm in FWHM. Cosmic rays were identified and removed from the datacubes using Renishaw's WiRE software. Peak parameters were extracted from Lorentzian fits, after a quadratic background was subtracted.

**Data clustering**

The GMM and PCA in this work were performed in Python, using the tools available in the open source package scikit-learn.[32]

## ACKNOWLEDGEMENTS

Particular thanks to Sebastian Wood, Filipe Richheimer, Helen Shamdasani, Alex Dexter, Teresa Murta and Spencer Thomas for their helpful discussion and feedback.

This project has received funding from the European Union's Horizon 2020 Research and Innovation program under grant agreement GrapheneCore3, number 881603. The work has also been financially supported by the Department for Business, Energy and Industrial Strategy though NMS funding (2D Materials Cross-team project). Additionally, this work was supported by JSPS KAKENHI grant number JP18H03864, and JST CREST grant number JPMJCR18I1.

## AUTHOR CONTRIBUTIONS

T.V. performed the Raman measurements, wrote the analysis code, and wrote the manuscript. K.K. and H.A. fabricated the TBLG sample. O.K., V.A. and H.A. contributed to discussions. O.K. oversaw the project. All authors contributed to the manuscript.

## COMPETING INTERESTS

The authors declare no competing interests.

# Fast Raman analysis and classification of twisted bilayer graphene using data clustering – Supplementary figures

Tom Vincent[1,2], Kenji Kawahara[3], Vladimir Antonov[4,2], Hiroki Ago[3] and Olga Kazakova[1,*]
1. National Physical Laboratory, Hampton Road, Teddington TW11 0LW, UK
2. Department of Physics, Royal Holloway University of London, Egham TW20 0EX, UK
3. Global Innovation Center (GIC), Kyushu University, Fukuoka 816-8580, Japan
4. Skolkovo Institute of Science and Technology, Moscow, 143026, Russia

Email: olga.kazakova@npl.co.uk

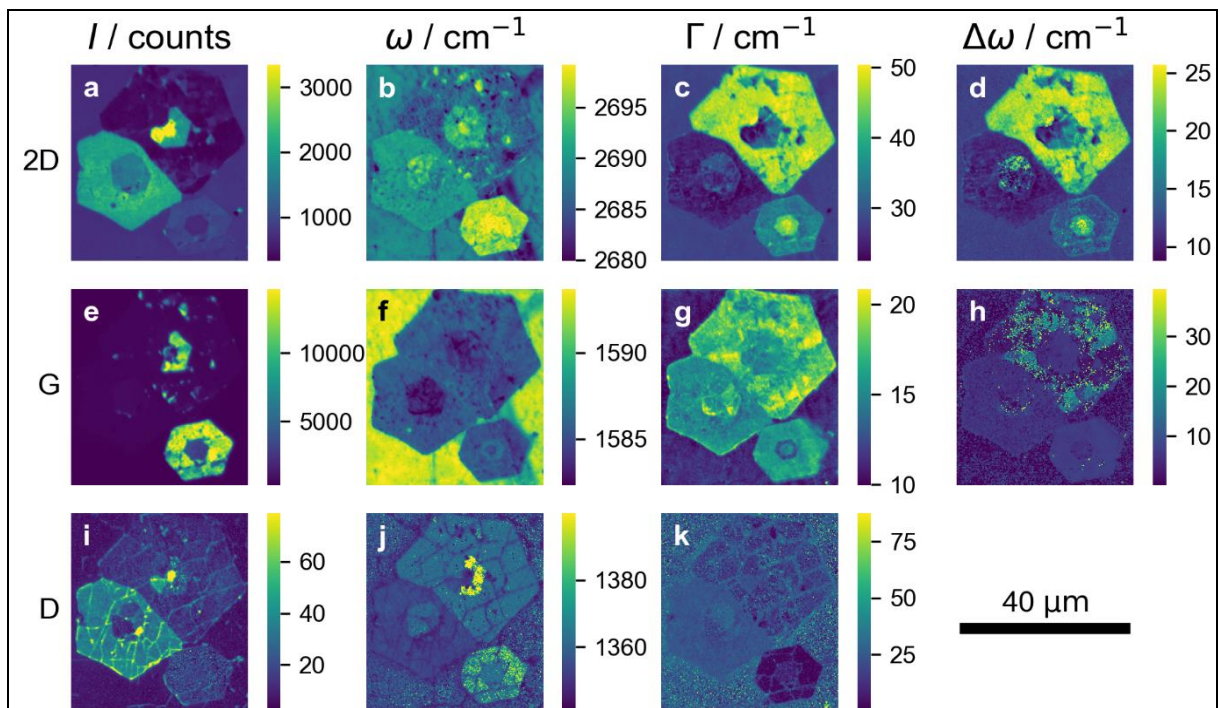

**Supplementary Fig. 1 Raman maps of region A. (a-k)** Maps of Raman peak intensity, position and FWHM for the 2D, G and D peaks, and peak splitting for the 2D and G peaks (**d** and **h**). All parameters were found from single Lorentzian fits, except for the splitting, which was found from separate double peak fits.

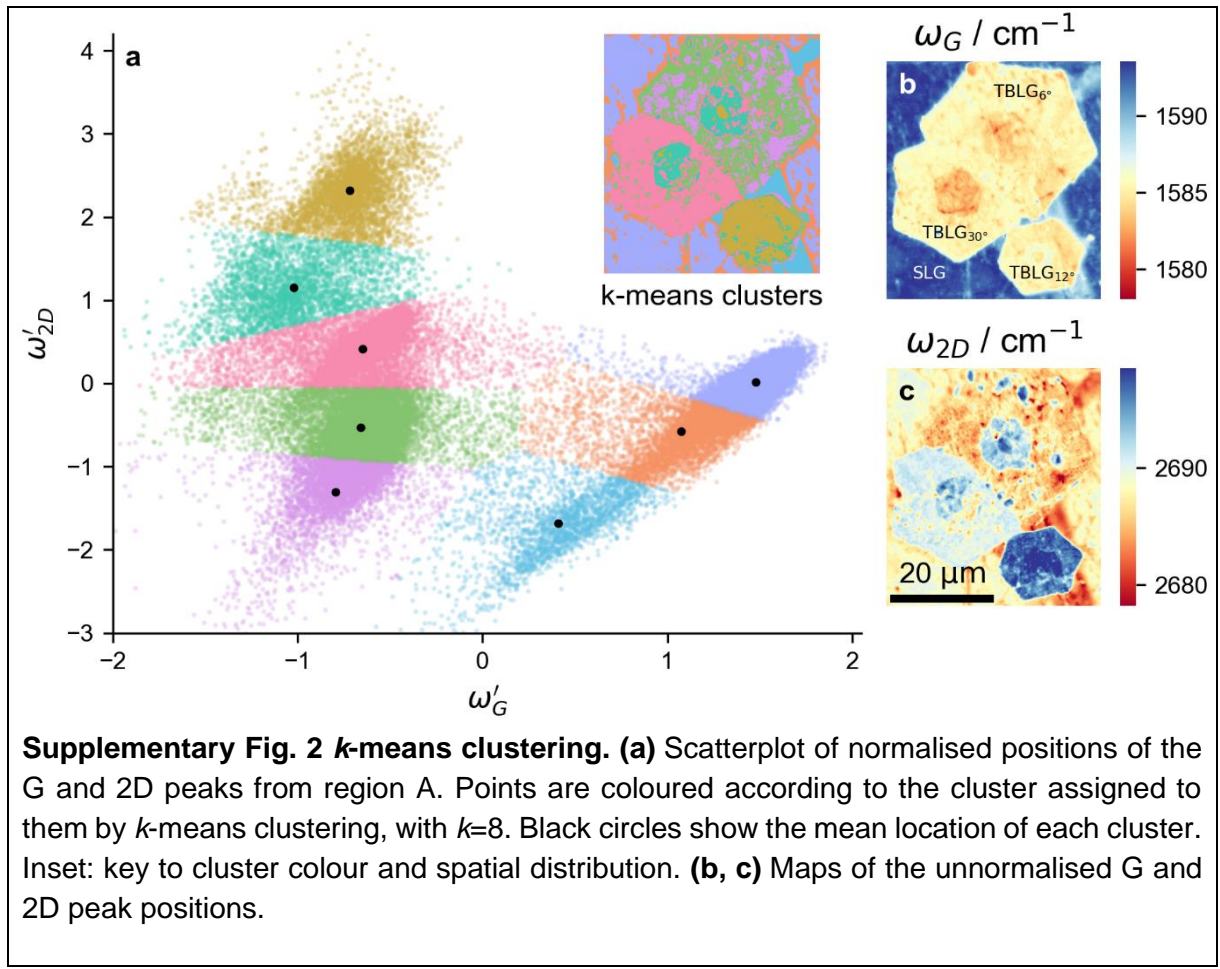

**Supplementary Fig. 2 *k*-means clustering. (a)** Scatterplot of normalised positions of the G and 2D peaks from region A. Points are coloured according to the cluster assigned to them by *k*-means clustering, with *k*=8. Black circles show the mean location of each cluster. Inset: key to cluster colour and spatial distribution. **(b, c)** Maps of the unnormalised G and 2D peak positions.

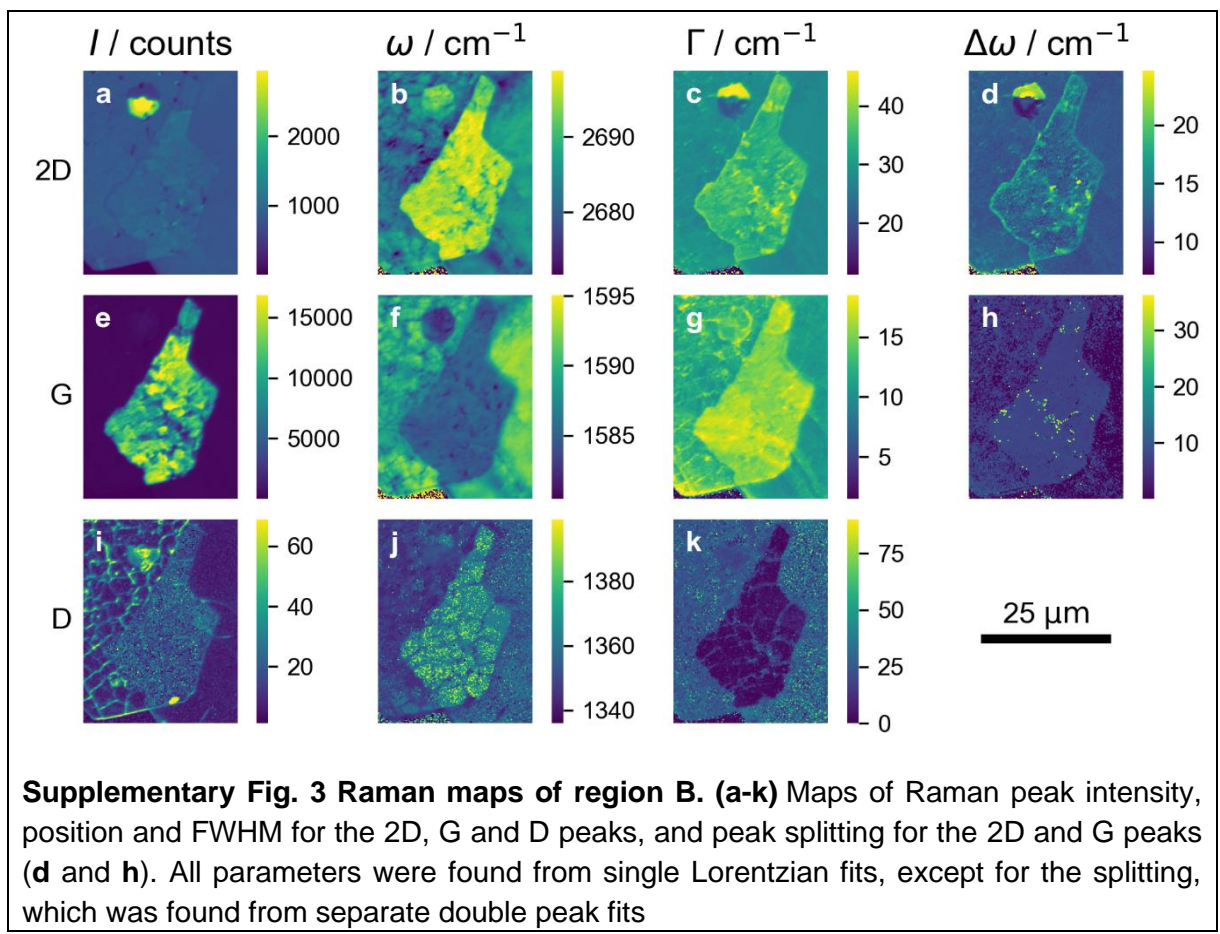

**Supplementary Fig. 3 Raman maps of region B. (a-k)** Maps of Raman peak intensity, position and FWHM for the 2D, G and D peaks, and peak splitting for the 2D and G peaks (**d** and **h**). All parameters were found from single Lorentzian fits, except for the splitting, which was found from separate double peak fits

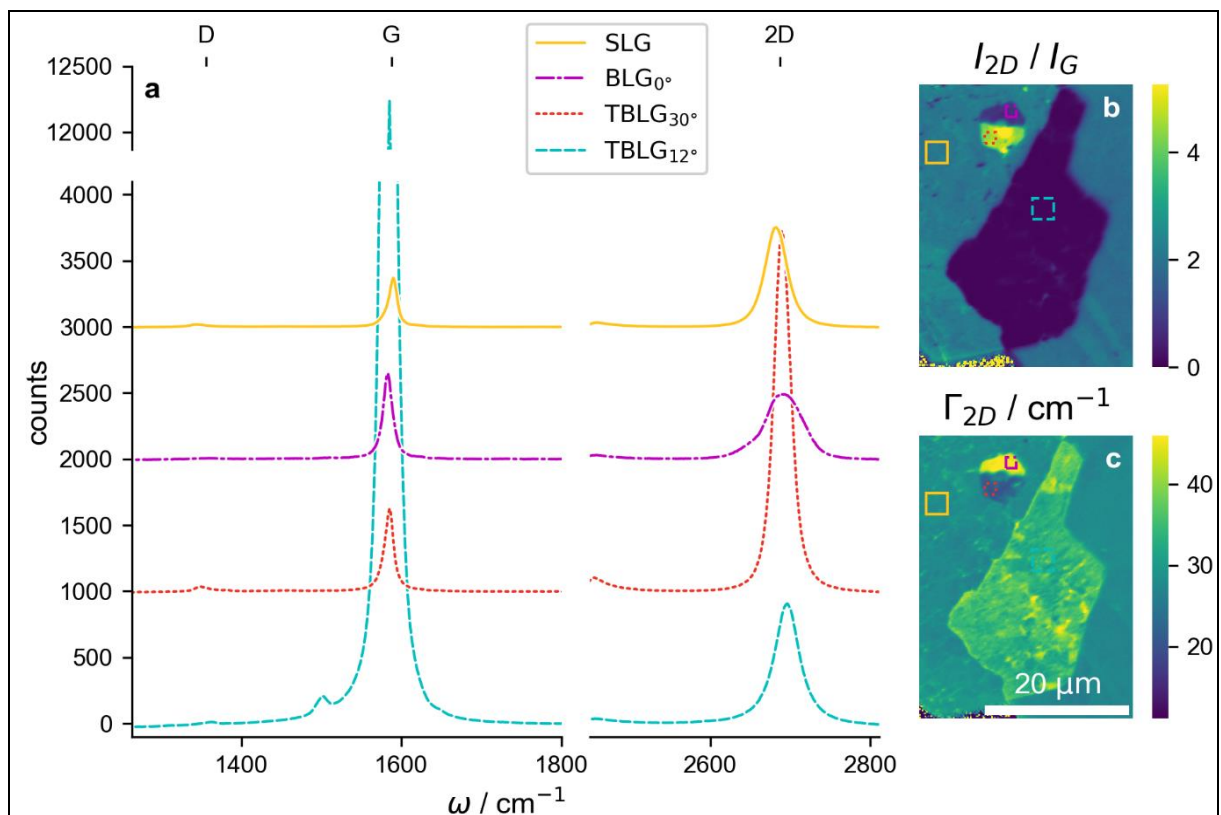

**Supplementary Fig. 4 Raman features of twisted bilayer graphene: region B.
(a)** Representative Raman spectra from key areas of the sample, showing large variation with twist angle. Spectra shown are averages of the spectra taken from within the squares of the corresponding colour and line style shown in **b** and **c**. It is worth noting that the 0° and 30° areas in this sample result from CVD island growth, not through stacking of different graphene batches. **(b, c)** Maps of the 2D to G peak intensity ratio and 2D peak FWHM.

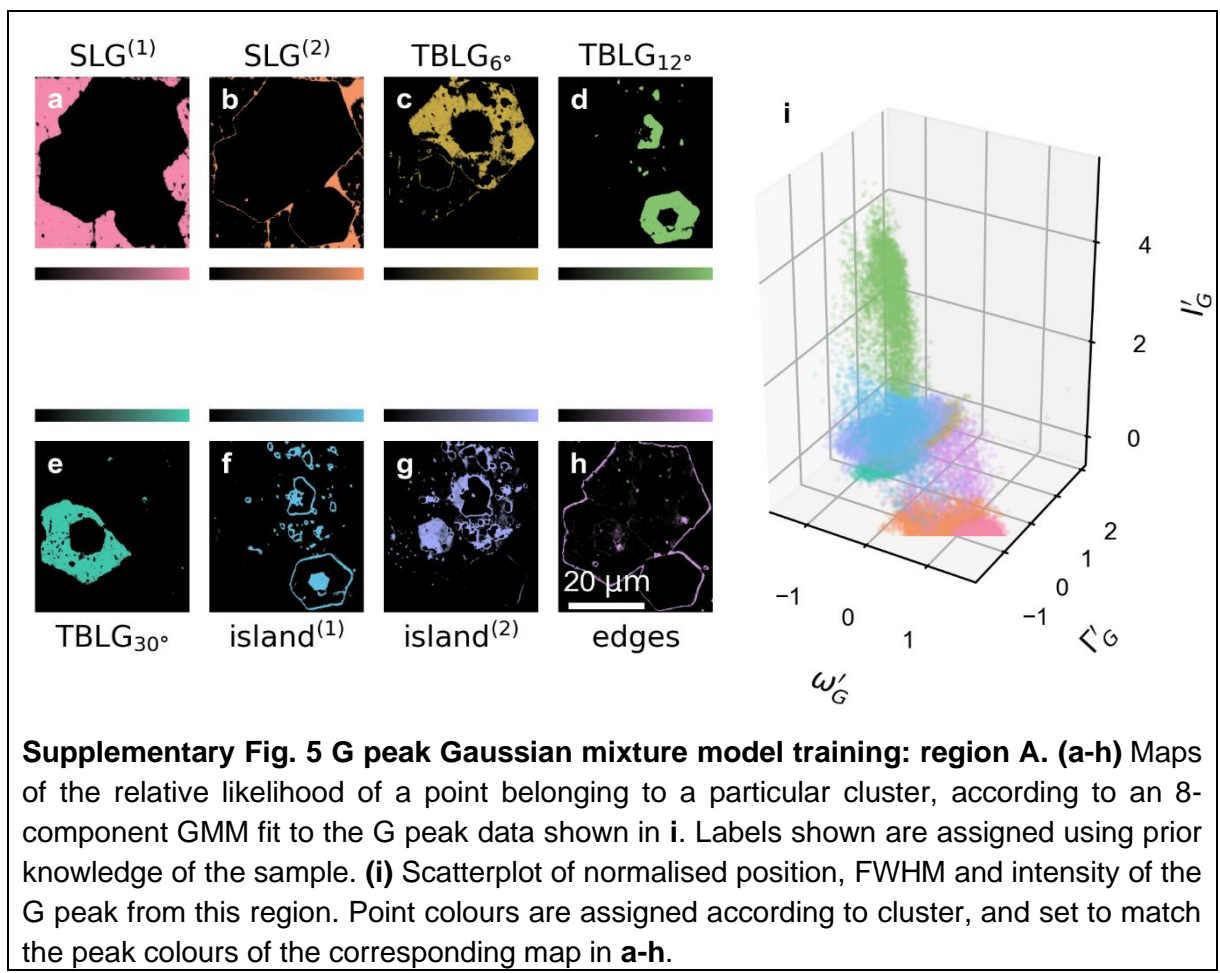

**Supplementary Fig. 5 G peak Gaussian mixture model training: region A. (a-h)** Maps of the relative likelihood of a point belonging to a particular cluster, according to an 8-component GMM fit to the G peak data shown in **i**. Labels shown are assigned using prior knowledge of the sample. **(i)** Scatterplot of normalised position, FWHM and intensity of the G peak from this region. Point colours are assigned according to cluster, and set to match the peak colours of the corresponding map in **a-h**.

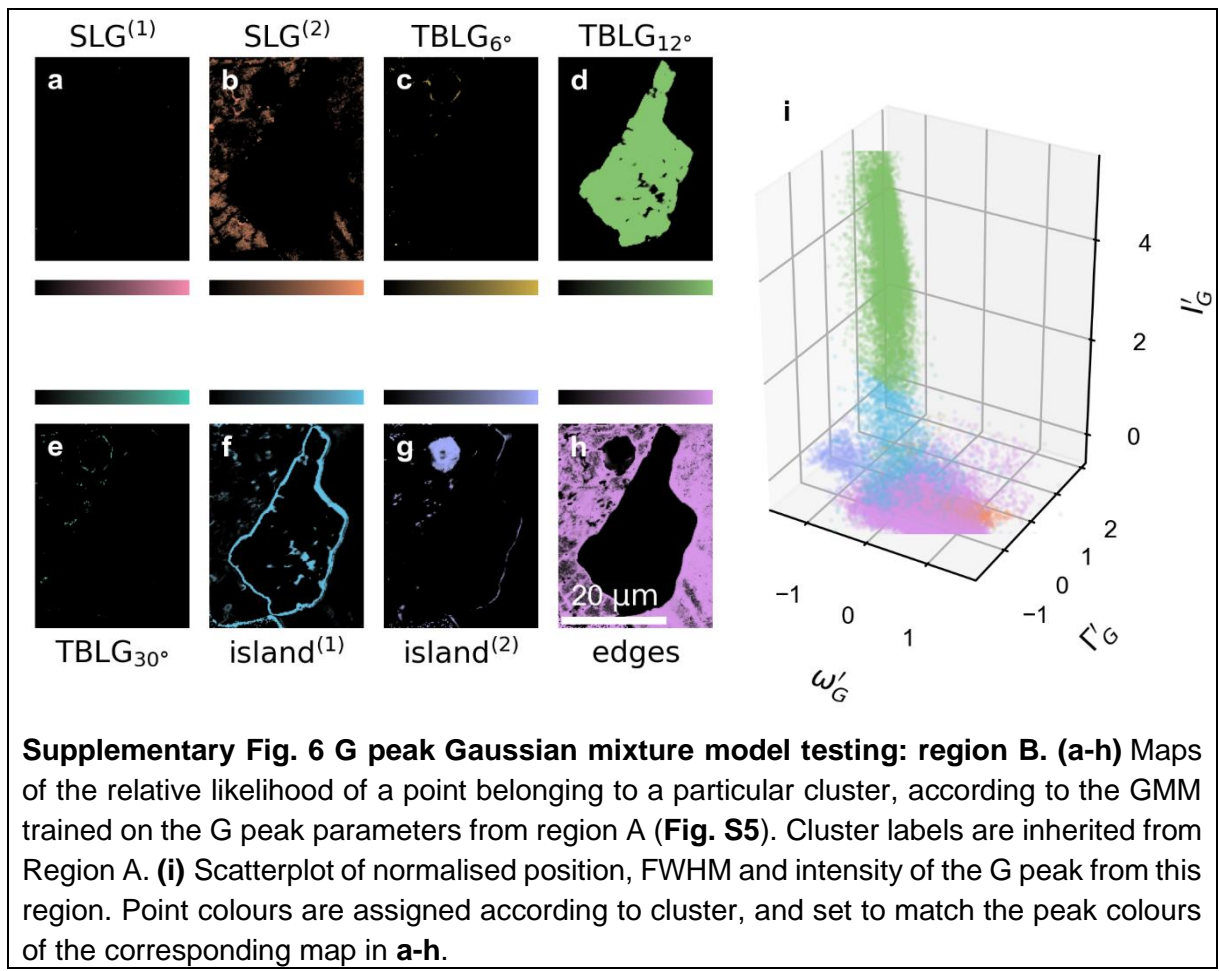

**Supplementary Fig. 6 G peak Gaussian mixture model testing: region B. (a-h)** Maps of the relative likelihood of a point belonging to a particular cluster, according to the GMM trained on the G peak parameters from region A (**Fig. S5**). Cluster labels are inherited from Region A. **(i)** Scatterplot of normalised position, FWHM and intensity of the G peak from this region. Point colours are assigned according to cluster, and set to match the peak colours of the corresponding map in **a-h**.

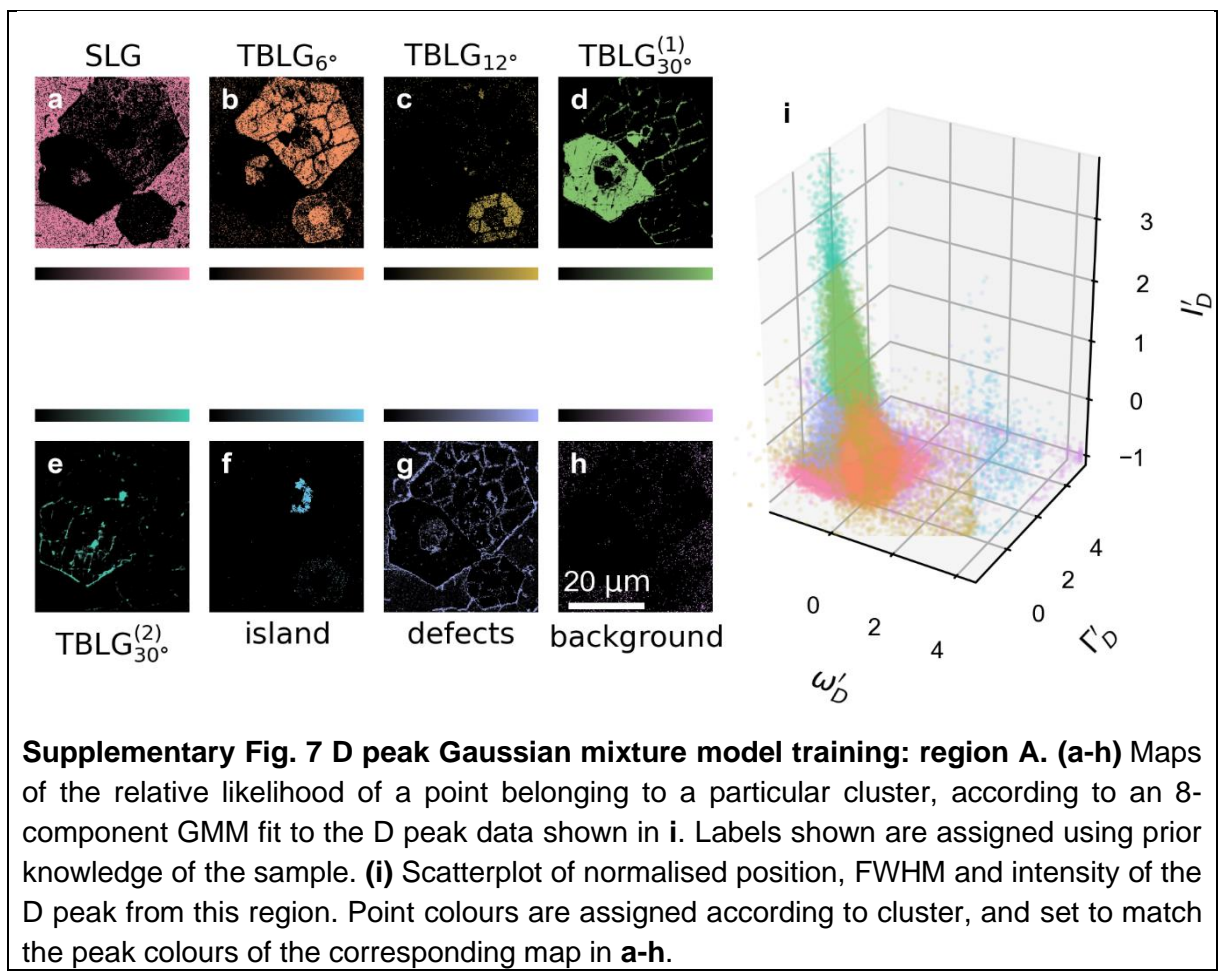

**Supplementary Fig. 7 D peak Gaussian mixture model training: region A. (a-h)** Maps of the relative likelihood of a point belonging to a particular cluster, according to an 8-component GMM fit to the D peak data shown in **i**. Labels shown are assigned using prior knowledge of the sample. **(i)** Scatterplot of normalised position, FWHM and intensity of the D peak from this region. Point colours are assigned according to cluster, and set to match the peak colours of the corresponding map in **a-h**.

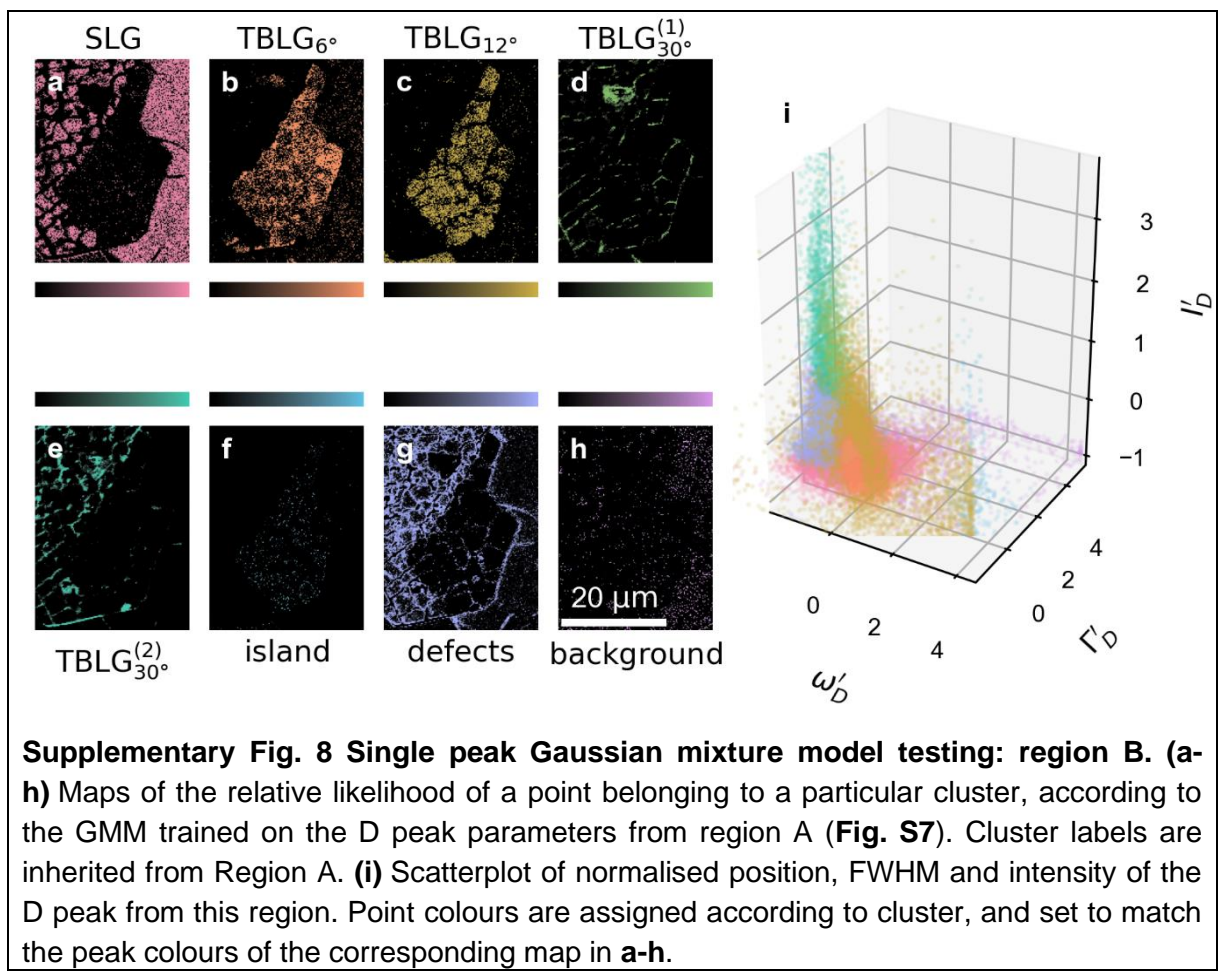

**Supplementary Fig. 8 Single peak Gaussian mixture model testing: region B. (a-h)** Maps of the relative likelihood of a point belonging to a particular cluster, according to the GMM trained on the D peak parameters from region A (**Fig. S7**). Cluster labels are inherited from Region A. **(i)** Scatterplot of normalised position, FWHM and intensity of the D peak from this region. Point colours are assigned according to cluster, and set to match the peak colours of the corresponding map in **a-h**.

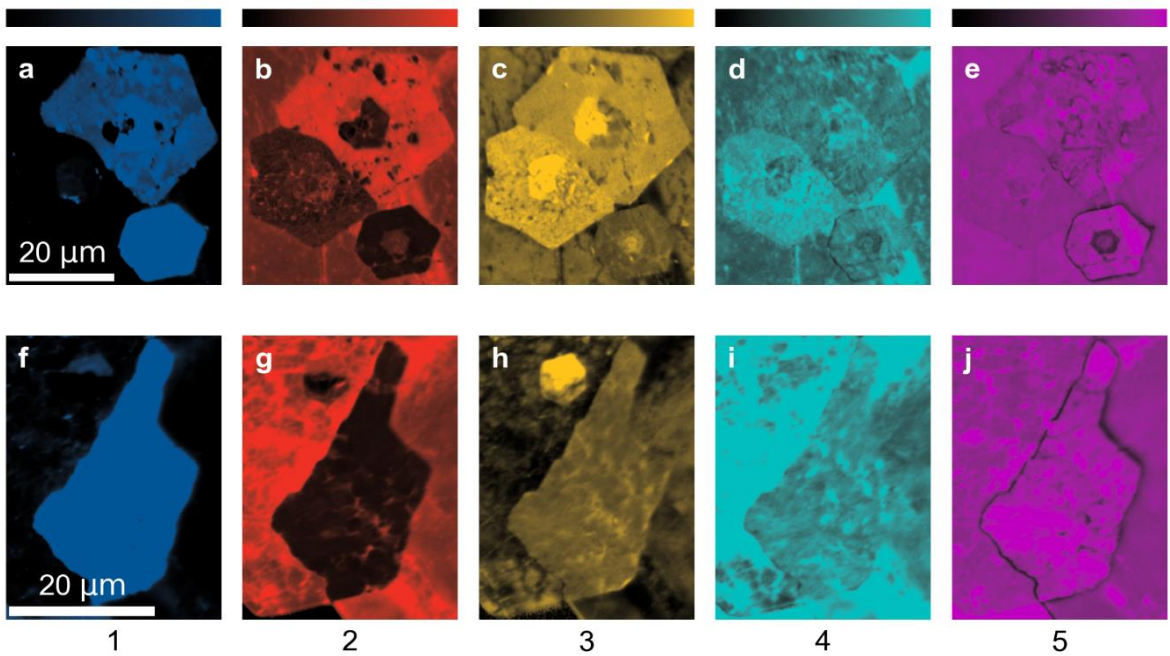

**Supplementary Fig. 9 Spatial distribution of principal components. (a-j)** Maps showing the relative intensity of the 5 selected principal components (shown in **Fig. 6**) in regions A and B.

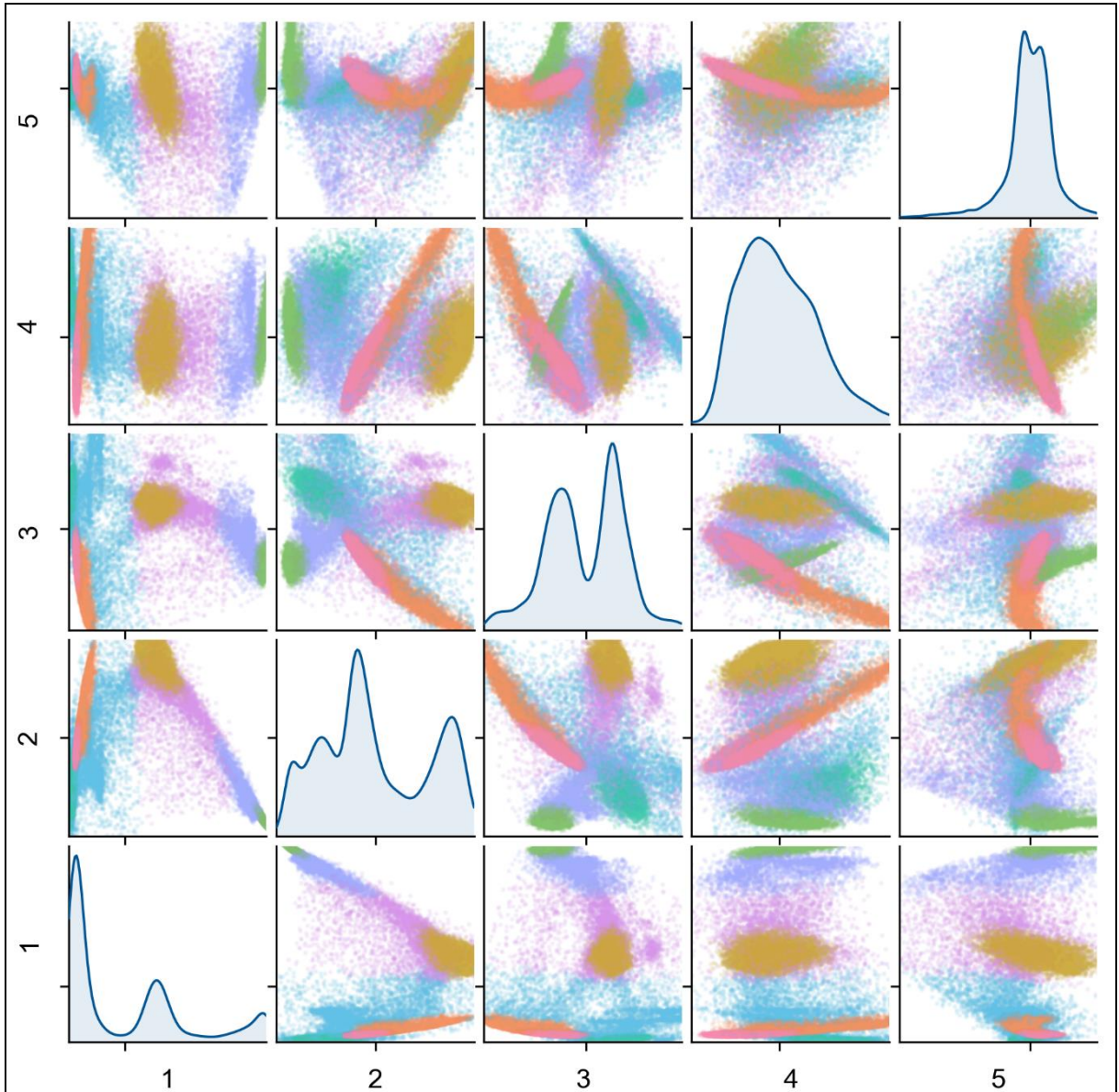

**Supplementary Fig. 10 Pairwise comparison of principal components: region A.** Scatterplots showing the relative distribution between each combination of the 5 selected principal components (shown in **Fig. 6**). Points are coloured by the cluster assigned by the full spectrum GMM, as shown in (**Figs. 7** and **8**). Data is in arbitrary units and axis ticks indicate the zero position. Plots along the central diagonal show the density distribution of each individual component.

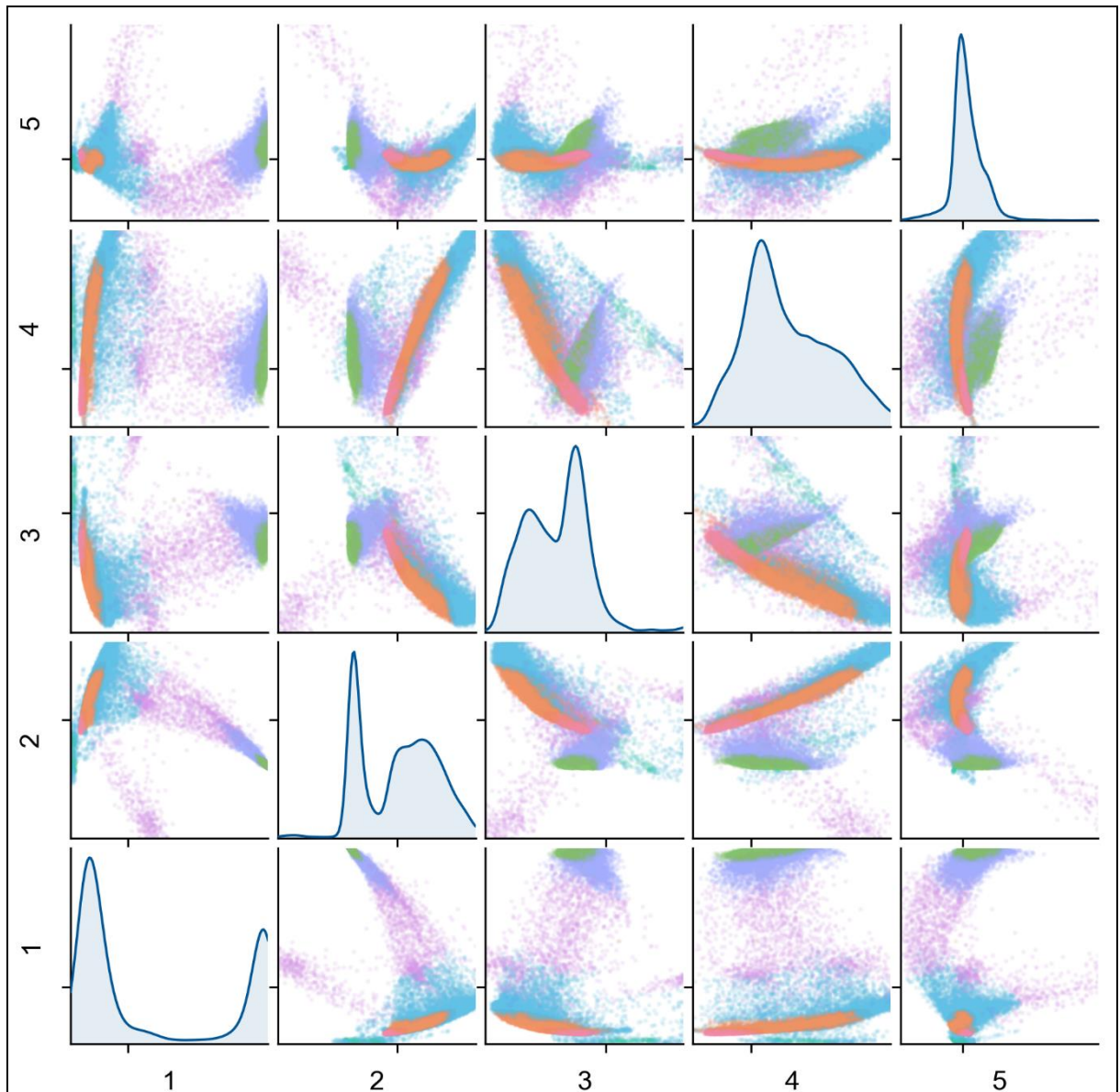

**Supplementary Fig. 11 Pairwise comparison of principal components: region B.** Scatterplots showing the relative distribution between each combination of the 5 selected principal components (shown in **Fig. 6**). Points are coloured by the cluster assigned by the full spectrum GMM, as shown in (**Figs. 7** and **8**). Data is in arbitrary units and axis ticks indicate the zero position. Plots along the central diagonal show the density distribution of each individual component.

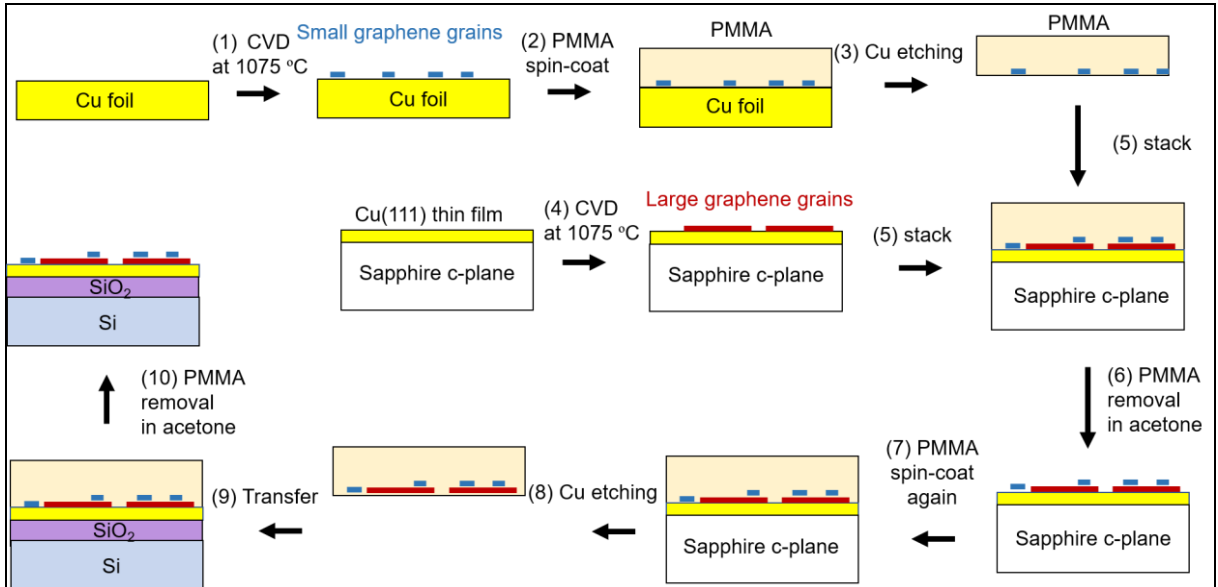

**Supplementary Fig. 12 Detailed transfer process of artificially stacked TBLG grains.** Images showing the step-by-step fabrication process for the TBLG sample used in this work. The removal and reapplication of PMMA in steps 6 and 7 was needed because the PMMA layer applied in step 2 was weakened during the Cu etching in step 3. The CVD growth in step 4 followed the process detailed in [ref. 31].